\documentclass[12pt]{article}
\usepackage{graphicx,color,epsfig}
\definecolor{pink}{rgb}{1.0,0.,1.0}

\begin{document}

\title{A Monte Carlo Generator for Full Simulation of $e^+e^-
\rightarrow hadrons $ Cross Section Scan Experiment }
\author{ 
D. Zhang, G. Rong, J.C. Chen \\
 {\em Institute of High Energy Physics, Beijing 100049, China}}
\date{   }
\maketitle

\begin{abstract}
{A generator with  $\alpha^2$ order radiative correction  effects,
including both the initial state radiative corrections and the photon
vacuum polarization corrections, is built for the full simulation of
$e^+e^- \rightarrow hadrons $ in cross section scan experiment in the
quarkonium energy region. The contributions from the hadron production
structures including the resonances
of $1^{--}$ quarkonium families and the light hadrons spectrum 
below 2 GeV as well as the QED continuum hadron spectrum are all taken
into account in the event generation. It was employed successfully to
determine the detection efficiency for selection of the events of
$e^+e^-\rightarrow hadrons$ from the data taken in the energy region
from 3.650 GeV to 3.872 GeV covering both the $\psi(3686)$ and $\psi(3770)$
resonances in BES experiment. The generator reproduces the properties of
hadronic event production and inclusive decays well.}
\end{abstract}


\section{Introduction }
    $\psi(3770)$ is the charmonium resonance with the lowest mass above
open charm-pair threshold. It is thought to decay almost entirely 
to $D\bar{D}$~\cite{psi2_dd}.
However, it has been a puzzle for a long time that some measured values
of $D\bar{D}$ production cross section $\sigma_{D\bar{D}}$ at the peak of
$\psi(3770)$
resonance fail to fit in the measured values of the cross section
$\sigma_{\psi(3770)}$ for
$\psi(3770)$ production at the peak. The discrepancy between the
$\sigma_{\psi(3770)}$ and $\sigma_{D\bar{D}}$ was historically
found to be more than $30\%$~\cite{nondd}~\cite{R-Z-C}.
To understand this long-standing puzzle, it is crucial to measure 
the $\psi(3770)$ resonance parameters and the $D\bar{D}$
cross section at the resonance peak precisely
in the same experiment.
Experimentally, $\psi(3770)$ 
resonance parameters are extracted from fitting the observed hadronic 
cross sections in the energy range which covers
the $\psi(3770)$
resonance in cross section scan experiment. However,
since the $\psi(3686)$ resonance is closed to the $\psi(3770)$,
and the cross section of continuum QED production 
exceeds the contribution of $\psi(3770)$ resonance the heavy
overlaps between them would affect the cross section
configuration in this energy region. In order to accurately measure
the parameters of the resonances one had better simultaneously to
deal with both the resonances and the contribution of continuum
production based on the same data set.
So, the determinations
of the efficiencies for detecting the inclusive hadronic events
point by point is necessary in the
energy region covering several production processe.

The determinations of efficiencies for
detecting the $e^+e^- \rightarrow hadrons $ events
in such energy region would be more complicate than that
in the case for which only single narrow resonance has been concerned
in the data taking, in which the initial state radiation (ISR) effect and
the continuum contribution are negligible small, or in the case of the
simple R value measurement in the $e^+e^-$ annihilation in the energy
region far away from any resonances. 
Due to the ISR effects, the actual energy to produce
final products would drop to rather wide region below the collision 
energy by photon emissions from the beam particles.  
The detector response, the composition of the different physical
processes in the whole region and even
the event selection criteria all vary with the
energies and finally affect the
detection efficiency at the collision energy points. To obtain the
efficiencies at each of the different energy points reliably,
a generator to produce the full processes of $e^+e^-\rightarrow
hadrons $ is developed.

Sometimes for convenience the contribution from the vacuum
polarization can be turned off in the event generation.
The possible interferences between the continuum hadron production
and the inclusive electro-magnetic (EM) decay mode of narrow
resonances are considered in the generator and can be turned on/off too. 

\section{Method}

\subsection{Radiative correction in $e^+e^-$ annihilation}

\begin{figure}[htb]
\centerline{\epsfig{file=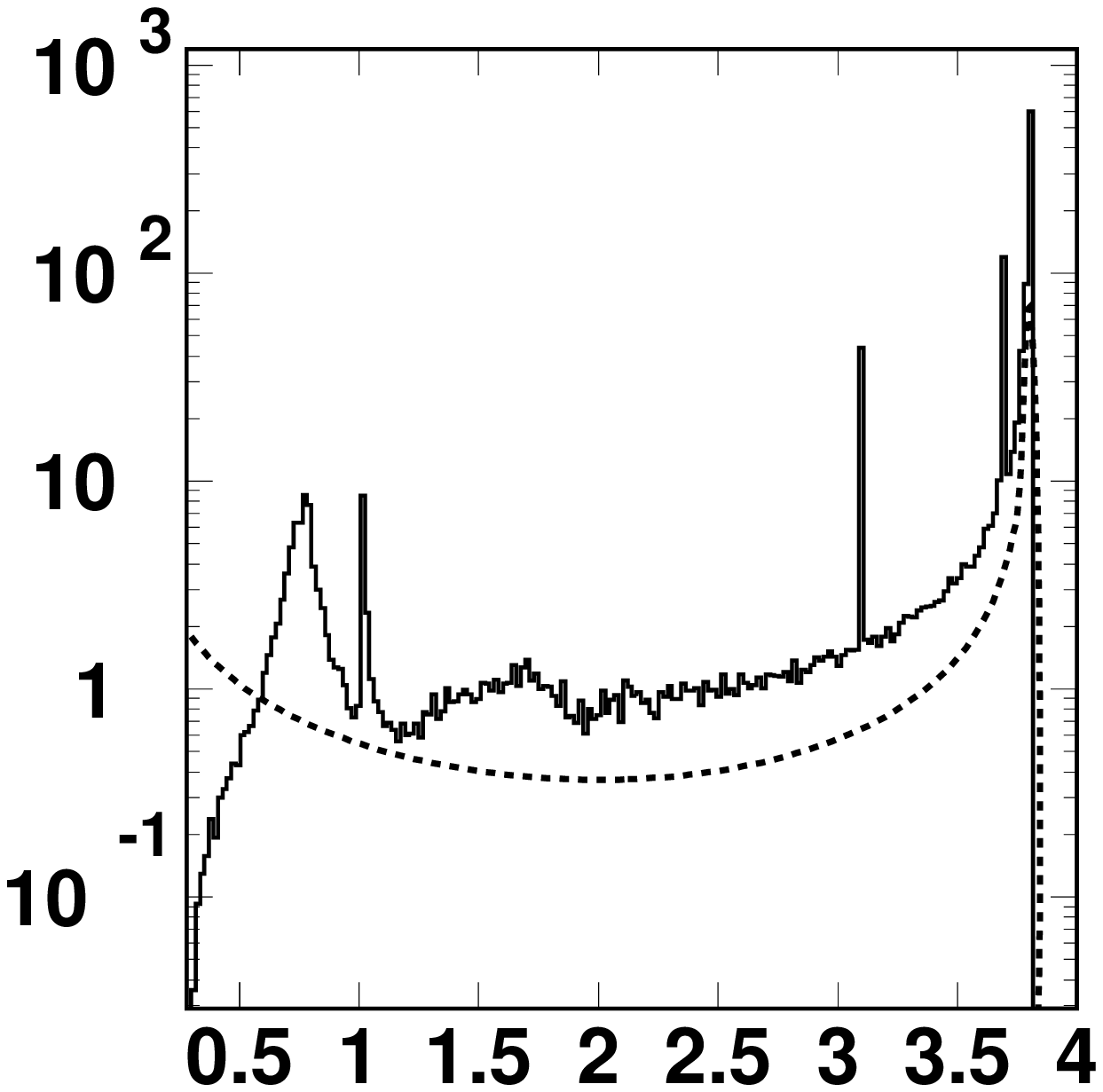,width=10cm,angle=0}}
\caption{The differential hadron production cross section due to ISR return
with respect to
the actual c.m. energy $\sqrt{s'}$, $d \sigma^{exp}(s)$/$d \sqrt{s'}$ in
the case that the nominal collision center-of-mass energy is set to 
$\sqrt{s}$ = 3.80 GeV. The dashed line shows the differential
cross section for $\mu^+\mu^-$ production.}
   \begin{picture}(5,5)
      \put(80,160){\rotatebox{90}{\large Density [nbar/GeV]}}
      \put(220,75){\large $\sqrt{s'}$ [GeV]}
\put(185.0,240.0){$\phi$}
\put(160.0,225.0){$\rho$}
\put(195.0,200.0){$\rho(1700)$}
\put(170.0,235.0){$\omega$}
\put(265.0,260.0){$J/\psi$}
\put(285.0,280.0){$\psi(2S)$}
\put(290.0,305.0){$\psi(3770)$}
\put(245.0,160.0){$\mu^+\mu^-$}
   \end{picture}
\label{resp}
\end{figure}
Due to ISR, the actual c.m. energy of the inclusive hadron system
produced in the $e^+e^-$ annihilation is
$\sqrt{s'}=\sqrt{s(1-x)}$, where $\sqrt{s}$ is the the collision
center-of-mass energy, and $x (1 > x\geq{0}) $ is a parameter relating to
the total energy of the emitted photons~\cite{isr}.

Fig.~\ref{resp} shows the differential cross section of the
hadron production with respect to $\sqrt{s'}=\sqrt{s(1-x)}$
for the case of that the nominal center-of-mass energy is set
to 3.80 GeV. 
To $\alpha^2$ order radiative correction in $e^+e^-$
annihilation,
the differential cross section of hadron production with
respect to the hard photon energy parameter $x$, instead
of $\sqrt{s'}$
can be written as
\begin{equation}
\frac{d \sigma^{exp}(s)}{dx}=
\frac{\sigma^{0}[s(1-x)]}{|1-\Pi[s(1-x)]|^{2}}F(x,s),
\label{equ_1}
\end{equation}
where $F(x,s)$ is a sampling function
for the radiative photon energy fraction $x$. Neglecting
the $Z^0$ exchange contributions
in the structure function approach introduced by Kuraev and Fadin~\cite{isr},
it can be written as
\begin{equation}
F(x,s)=\beta x^{\beta-1}\delta^{V+S}+\delta^{H},
\label{equ_F}
\end{equation}
in which $\beta$ is the electron equivalent radiator thickness,
$$\beta=\frac{2\alpha}{\pi}(L-1)$$
and
$$ L={\rm ln}\frac{s}{m_{e}^{2}}$$
where $m_e$ is the mass of electron, $\alpha$ is the 
fine structure constant, and
\begin{equation}
\delta^{V+S}=1+\frac{3}{4}\beta+\frac{\alpha}{\pi}(\frac{\pi^2}{3}-
\frac{1}{2})+\frac{\beta^2}{24}(\frac{37}{4}-\frac{L}{3}-2\pi^2),
\label{equ_4}
\end{equation}
\begin{equation}
\delta^{H}=\delta_1^H+\delta_2^H,
\label{equ_5}
\end{equation}
in which
$$\delta_1^{H}=-\beta(1-\frac{x}{2})$$
and
$$\delta_2^{H}=\frac{1}{8}\beta^2[4(2-x){\rm ln}\frac{1}{x}-
\frac{1+3(1-x)^2}{x}{\rm ln}(1-x)-6+x].$$
In Eq.~(\ref{equ_1}), 
1/$|1-\Pi[s(1-x)]|^{2}$ is the vacuum polarization 
correction factor including both  
the leptonic and the hadronic terms i.e.
\begin{equation}
\Pi(s)=\Pi_{h}(s)+\Pi_{l}(s)
\label{equ_Pi}
\end{equation}
in which
$$\Pi_{h}(s)=\frac{s}{4\pi^{2}\alpha}[PV\int\frac{\sigma^{0}(s')}
{s-s'}ds'-i\pi\sigma^{0}(s)]$$
and
$$\Pi_{l}(s)=\frac{\alpha}{\pi}f(\xi),$$
where
$$f(\xi)=-\frac{5}{9}-\frac{\xi}{3}+\frac{\sqrt{1-\xi}(2+\xi)}
{6}{\rm ln}[\frac{1+\sqrt{1-\xi}}{1-\sqrt{1-\xi}}]~~(\xi \leq{1})$$
and
$$f(\xi)=-\frac{5}{9}-\frac{\xi}{3}+\frac{\sqrt{1-\xi}(2+\xi)}{3}
{\rm tan^{-1}}\frac{1}
{\sqrt{\xi-1}}~~(\xi >1),$$
in which $\xi=\frac{4m^{2}_{l}}{s}$ where $ m_{l}$ is the mass of
lepton $l (l = e, \mu, \tau) $.
In Eq.~(\ref{equ_1}),
\begin{equation}
\sigma^{0}[s(1-x)]= \frac{4\pi\alpha^2}{3s(1-x)}R+
\sum_{i}\sigma_{res,i}^{0}[s(1-x)]
\label{equ_Born1}
\end{equation}
is the total lowest order cross-section at $\sqrt{s'}=\sqrt{s(1-x)}$ in 
which $R$ is the ratio of continuum hadron cross section to the
cross section of $e^+e^-\rightarrow \mu^+\mu^- $, and
\begin{equation}
\sigma_{res,i}^{0}[s(1-x)]= 
\frac{12\pi\Gamma_{h,i}\Gamma_{ee,i}^{0}}
{[s(1-x)-(M_i^0)^2]^2+(M_i^0)^2\Gamma_i^2}
\label{equ_Born2}
\end{equation}
is the lowest order cross section of the $i$th resonance, where $\Gamma_{ee,i}^0$,
$M_i^0$, $\Gamma_{h,i}$ and $\Gamma_i$ are the lowest order leptonic
width, the lowest order mass, the hadronic width and the total width of
the $i$th resonance, and '$i$' denote the $i$th. resonance of the $1^{--}$ quarkonium
states and the states with mass below 2 GeV/$c^2$. 
For wide resonances the widths  $\Gamma_i$'s and $\Gamma_{h,i}$ vary with
energy depending on their decay products.

\begin{figure}[htb]
\centerline{\epsfig{file=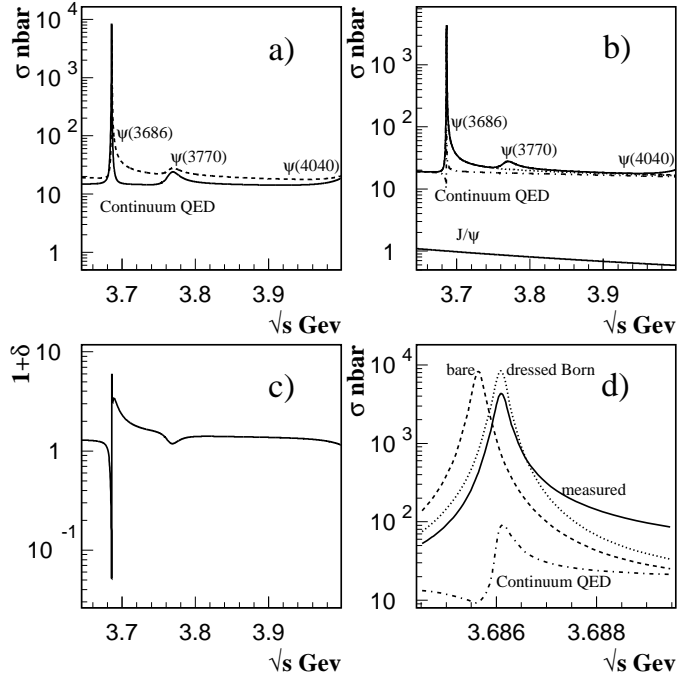,width=10cm,angle=0}}
\caption{The cross sections of $e^+e^-$ annihilation into inclusive
hadron in the energy region from 3.65 GeV to 4.0 GeV. a) shows the total
lowest order cross section (solid line) and
the expected total cross section in the region, produced by
the generator with null collision energy spread setting (dashed line);
b) demonstrates the respective contributions of the expected
cross sections of the five processes mentioned in the text; 
c) represents
the radiative correction factor $1+\delta$;
and d) gives some detail information about the radiative corrections 
at the position of the sharp resonance $\psi(3686)$ in the tiny energy
window from 3.6845 GeV to 3.6895 GeV.
The EM interference effect between $\psi(3686)$ and continuum QED is
turned off in the event generation.}
\label{xborn}
\end{figure}

In Fig.~\ref{xborn} a), the solid line shows the total lowest order cross 
section for inclusive hadronic event production in the region from
3.65 GeV to 4.0 GeV including the five processes:
$e^+e^-\rightarrow J/\psi$, 
$e^+e^-\rightarrow \psi(3686)$, 
$e^+e^-\rightarrow \psi(3770)$,
$e^+e^-\rightarrow \psi(4040)$ and 
$e^+e^-\rightarrow \gamma^*\rightarrow hadrons$;
while the dashed line shows the expected total cross section with the
radiative corrections for the inclusive hadron production.
These cross sections are generated with the generator.
In Fig.~\ref{xborn} b), the contributions of
the radiatively corrected cross sections of the five processes
are demonstrated, respectively.
The energy spread for generating the events is set to be zero.

The ratio of the expected observation cross section to the lowest order
cross section, see the two lines in Fig.~\ref{xborn} a) is the
radiative correction
factor for the inclusive $e^+ e^- \rightarrow hadrons $ process
\begin{equation}
     1+\delta=\frac{\sigma^{exp}(s)}{\sigma^0(s)},
\end{equation}
which is shown in Fig~\ref{xborn} c).

From Fig.~\ref{xborn} c) one can see that the radiative
correction effects vary rather large around the resonance.
Except for the ISR corrections, under the narrow resonance peak 
the large real and imaginary part of vacuum polarization correction would
shift the lowest order resonance peak position $M^0$ up by an amount of
\begin{equation}
   \Delta M =\frac{\frac{3}{2\alpha}\Gamma_{ee}^0
  +\frac{\alpha}{6}R\Gamma}{1-Re\Pi^{\prime}[(M^0)^2]},
\label{m_shift}
\end{equation}
where $R$ denotes the $R$ value for the continuum hadron production and
$\Pi^{\prime}[(M^0)^2]$ denotes
the vacuum polarization function from all the contributions of leptons
and hadrons except for the contribution from the resonance itself at the
position $s=(M^0)^2$.
And
\begin{equation}
    M=M^0+\Delta M
\label{m_dress}
\end{equation}
is the observable peak position of the resonance related to the single photon
exchange processes. One can also dress
the lowest order leptonic width $\Gamma_{ee}^0$ to the physical one $\Gamma_{ee}$
in the same way, by
\begin{equation}
  \Gamma_{ee}=\frac{\Gamma_{ee}^0}{|1-Re\Pi^{\prime}[(M^0)^2]|^2},
\end{equation}
for calculation of ${d \sigma^{exp}(s)}/{dx}$ given in Eq.~(\ref{equ_1}).
The change of the total width $\Gamma$ due to the vacuum polarization
is negligibly small.
The dashed line and dotted line in Fig.~\ref{xborn} d) show the
split of $\psi(3686)$ peak due to the EM vacuum polarization
given in Eq.~(\ref{m_shift}). The label 'bare' in Fig.~\ref{xborn} d)
indicate the lowest order peak position which can be seen in the
processes in which there are no EM effects infected. 

\begin{figure}[htp]
\centerline{\epsfig{file=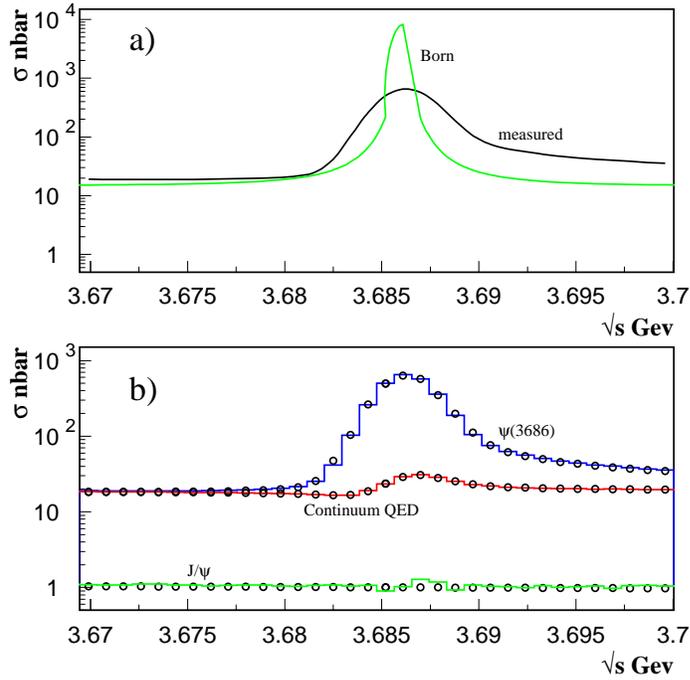,width=10cm,angle=0}}
\caption{The simulation of a fine cross section scan in $\psi(3686)$
resonance region where the energy spread is assumed to be
1.35 MeV. In a), the expected observation cross section is shown
by the solid line, while 
the green line shows the lowest cross section;
and in b) the contributions from
the three observation cross sections, i.e. the $J/\psi$ tail, the
$\psi(3686)$ and the continuum QED process, are demonstrated. The
circle marks on the lines in b) represent the integrated results
of Eq.~(\ref{equ_1}) for the cross sections of the three processes.}
\label{psip}
\end{figure}

The vacuum polarization effects also cause the flat continuum QED
contribution
vibration under the sharp resonance, which is shown by
dashed dotted line in Fig.~\ref{xborn} d) and also in Fig.~\ref{psip} b)
by the red line
The solid line in Fig.~\ref{xborn} d) represents the expected
total cross section in the energy region.

Fig.~\ref{psip} a) shows the different behavior of the lowest order
cross section and the expected observable cross section due to radiative
effects and with finite collision energy spread (1.35 MeV) generated
with the generator in the $\psi(3686)$ resonance region.
While Fig.~\ref{psip} b) shows the contributions of the observation cross
sections of three relative processes in the energy region, the tail of
$J/\psi$, the $\psi(3686)$ resonance and the continuum QED, in which
the vibration of the continuum QED line shape caused by the vacuum
polarization corrections is still seen clearly in the finite energy
spread case (the red line).
The finite energy spread does drop the sharp resonance peak significantly
but does not eliminate the vibration
of continuum QED line shape as shown in the figure.
If the data handling, the event selection, the determination
of the detection efficiencies and the physical considerations
of the hadron spectrum structure are all
treated properly and correctly in the cross section scan experiments,
one can expect that the vibration effects of continuum QED line shape
due to vacuum polarization lying
on the narrow resonance region would be observable.  

The circle marks on the lines in Fig.~\ref{psip} b)
are the integrated numerical results of Eq.~(\ref{equ_1}) for the
corresponding cross sections.
The accuracies of the simulated cross sections by the generator are quite
good. 

\subsection{Sub-generators}
From Fig.~\ref{resp}, one can see that
the hadronic events
not only be produced at the $\sqrt{s}$, but also produced in the full
energy range from $\sqrt{s}$ to 0.28 GeV, the two
pion production threshold.
To correctly simulate the full process produced in $e^+e^-$
annihilation,
the generator should be a mixture of sub-generators in which
each single generator simulates a special processes and runs
together with others, driving according to the ISR return energy
spectrum weighted by the lowest order cross sections of the processes
involved in the energy region. Up to the charmonium system energy region,
these physics processes include one photon continuum hadron
production,
$e^+e^-\rightarrow J/\psi, \psi(3686)$, which decay to hadronic final
states inclusively,
the $e^+e^-\rightarrow
\psi(3770),
\psi(4040)$ and $\psi(4160)$ etc. which go to the hadron final
states through the mediate charmed meson pairs inclusively.
The production of $1^{--}$ resonances at lower energy
range are also included, such as the $\rho(1700)$, $\phi(1680)$,
$\omega(1650)$, $\rho(1450)$, $\omega(1420)$, $\phi(1020)$,
$\omega(780)$ and $\rho(770)$ etc.. Those states are arranged
respectively to 
go to various final states such as $ \pi\pi, K \overline{K},
K \overline{K}\pi, 4\pi, 5\pi, 6\pi$.... exclusively with their
corresponding cross sections and branching fractions. 
The events of $e^+e^-\rightarrow \gamma^*\rightarrow
hadrons$ and the inclusive hadron decays of
charmonium resonance $\psi(3686)$ and $J/\psi$ 
are generated by their respective inclusive LUND-type generators
~\cite{lund}~\cite{lund_charm}.
While the events which decay from other resonances such as the $\psi(3770)$,
$\phi(1680)$, $\phi(1020)$, $\rho(1700)$, $\rho(1450)$, $\rho(770)$
and $\omega(782)$ etc. to all possible final
states according to the known decay modes and branching
fractions~\cite{PDG}, can be generated by corresponding
sub-generators conveniently.
The Lorentz boost of the hadron system recoiling against the
radiative photons are planted into the sub-generator's operation.
A normal distribution random
number generator has been employed to simulate the effects of energy spread
at the collision energy $\sqrt{s}$. 

\subsection{Sampling Technique}
    From Eq.~(\ref{equ_Born2}), Eq.~(\ref{equ_F}) and Fig.~\ref{resp}
one can see that there are many sharp peaks, such as the $J/\psi$,
$\psi(3686)$ peaks (see Fig.~\ref{resp} and Fig.~\ref{xborn}) and
the sharp photon ISR
spectrum at the singularity when $ x\rightarrow 0$ (equivalent to
$\sqrt{s'}\rightarrow $ 3.8 GeV in Fig.~\ref{resp}), which enter into the
probability distribution function given in Eq.~(\ref{equ_1}).
All of those sharp structures may slow greatly the Mont Carlo
sampling procedure and affect the accuracy of the event generations,
especially for the accuracies of the shapes of the sharp variation lines
which are extremely important for the determination of the detection
efficiencies.
Sometimes people can eliminate the singularities
by Monte Carlo inverse transformation techniques, but it is almost
impossible to eliminate all those singularities simultaneously by
an inverse transformation function. A special importance sampling
method with multi-section scheme treatment has been developed in this work.
One divides the whole of ISR return energy region into several sections
to guarantee that for every section there is only one or less peak of the
distribution function given by Eq.~(\ref{equ_1}).
Three kinds of sections are defined: 1) the resonance
section in which there is only one sharp resonance like $J/\psi$ or
$\psi(3686)$;
2) the slow varying sections in which there is no sharp resonance
or any singularity; and 3)
the section in which the ISR spectrum has the singularity i.e. the
region which directly touches to the
collision energy point $x\rightarrow 0, (s'\rightarrow s) $, for
which the section range is
variable and depending on the position of the nearest sharp resonance
beneath the collision energy $ \sqrt {s} $.
When the collision energy with its energy spread range close to
the narrow resonance or overlap each other, the width of the
ISR singularity section $x_0$ ($x_0 \geq{x} \geq{0}$) should  
ensure that both of the variation of the lowest order cross section in the
singularity section region ($x_0 \geq{x} \geq{0}$) and the variation of
the ISR spectrum in the touched sharp resonance section should be comparably
moderate enough.

The sharp resonance peak in the resonance section can be
eliminated easily by the inverse function transformation,
defining the new  variant $ \rho$ instead of $x$,
in Eq.~(\ref{equ_1})
\begin{equation}
  x=1-[M\cdot\Gamma\cdot{tan}(tan^{-1}(\frac{s(1-x^{up})-M^2}{M\cdot\Gamma}
    -\Omega\cdot\rho)+M^2]/s,
\end{equation}
in which
\begin{equation}
      \Omega=tan^{-1}\frac{s(1-x^{up})-M^2}{M\cdot\Gamma}-
             tan^{-1}\frac{s(1-x^{low})-M^2}{M\cdot\Gamma}.
\end{equation}
serves as the  probability normalization constant in the transformation,
and $x_{up}$ and $x_{low}$ are the up and low limits corresponding
to the resonance section. 
The $\rho$ is then the normalized random number taken uniformly
from 0 to 1, which keeps the event production probability, i.e. the
integrated area of Fig.~\ref{resp} unchanged in the resonace section
under the transformation.
This transformation keeps the exact
shape of the sharp resonance, but it blocks
generating the initial state photons individually. It can just output
the total energy and momentum of the emitted photons to serve for the
Lorentz boost of the recoiling hadronic system. For the purpose of
inclusive hadron measurement in $e^+e^-$ collision experiment this
simplification is feasible.
The price paid here is worthy for the precise measurement of the narrow
resonance parameters.

In this stage one can simply define the new normalized
sampling random number $\xi$ in (0,1) section instead of $x$
to eliminate the singularity of the exponential ISR
spectrum, the leading logarithm term of Eq.~(\ref{equ_F}) and
keep the event production probability unchanged,
\begin{equation}
    x=x_{0}\xi^{1/\beta},
\label{equ_x}
\end{equation}
where the $\beta$ is as defined in Eq.~(\ref{equ_F}),
and the section width $x_{0}$ here saves as the probability normalization
constant in the transformation.

After performing the importance sampling procedure for the multi-section
process according to the distribution of Eq.~(\ref{equ_1}) and the
normalized ratios of the lowest order cross sections of the processes
involved to drive the corresponding sub-generators the Mont Carlo events
will be then generated finally as mentioned in above section.

\begin{figure}[htbp]
\centerline{\epsfig{file=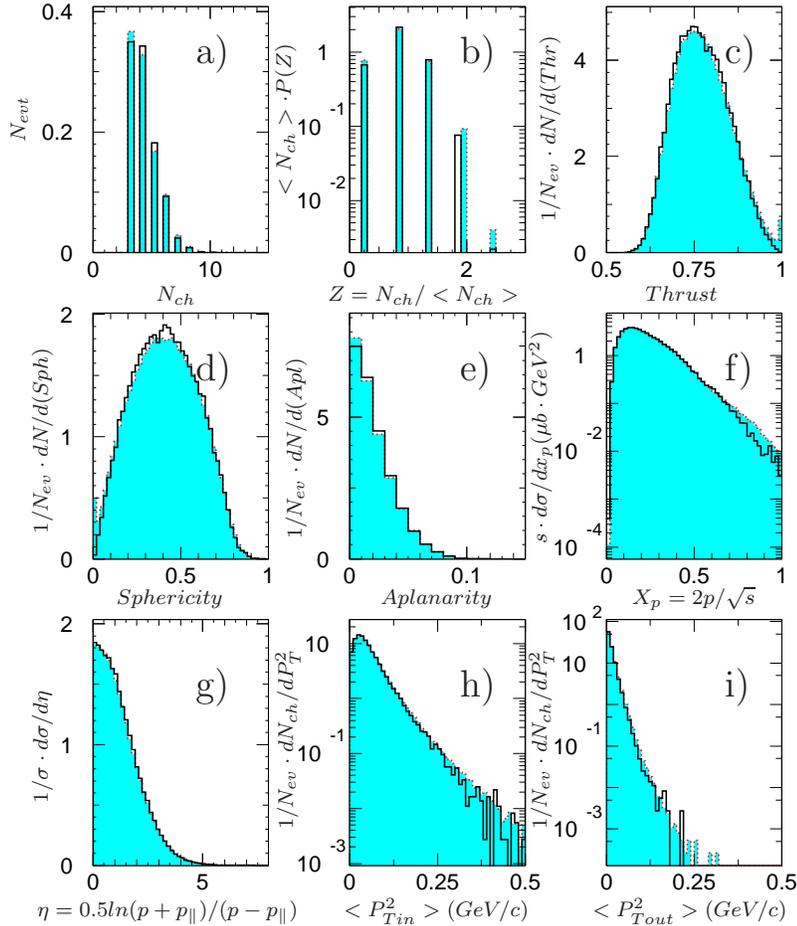,width=10cm,angle=0}}
\caption{The properties of hadronic events produced at $\sqrt{s}=3.773$ GeV. 
Shadows are the distribution of experimental data~\cite{chenjc_talk}, and lines 
are those generated with this generator;
a) the charged multiplicity; b) the $KNO$ scaling~\cite{KNO}; 
c) thrust~\cite{thrust}, d) sphericity~\cite{sph}, e) 
aplanarity~\cite{sph}
f) Feynman scaling variable, g) pseudo-rapidity, h) and i)
transverse momentum projection in and out the jet injection plane} 
\label{psi2}
   \begin{picture}(5,5)
      \put(150,410){\large a)}
      \put(250,410){\large b)}
      \put(350,410){\large c)}
      \put(150,290){\large d)}
      \put(250,290){\large e)}
      \put(350,290){\large f)}
      \put(150,170){\large g)}
      \put(250,170){\large h)}
      \put(350,170){\large i)}
      \put(80,380){\rotatebox{90} {\scriptsize $N_{evt}$}}
      \put(135,320){\scriptsize $N_{ch}$}
      \put(180,360){\rotatebox{90} {\scriptsize $<N_{ch}>\cdot P(Z)$}}
      \put(200,320){\scriptsize $Z=N_{ch}/<N_{ch}>$}
      \put(280,350){\rotatebox{90} {\scriptsize $1/N_{ev}\cdot dN/d(Thr)$}}
      \put(320,320){\scriptsize $Thrust$}
      \put(88,232){\rotatebox{90} {\scriptsize $1/N_{ev}\cdot dN/d(Sph)$}}
      \put(120,205){\scriptsize $Sphericity$}
      \put(184,230){\rotatebox{90} {\scriptsize $1/N_{ev}\cdot dN/d(Apl)$}}
      \put(220,205){\scriptsize $Aplanarity$}
      \put(276,228){\rotatebox{90} {\scriptsize $s\cdot d\sigma/dx_p(\mu b\cdot GeV^2)$}}
      \put(315,205){\scriptsize $X_p=2p/\sqrt{s}$}
      \put(88,130){\rotatebox{90} {\scriptsize $1/\sigma \cdot d\sigma/d\eta$}}
      \put(90,87){\scriptsize $\eta=0.5ln(p+p_{\parallel})/(p-p_{\parallel})$}
      \put(180,120){\rotatebox{90} {\scriptsize $1/N_{ev}\cdot dN_{ch}/dP^2_T$}}
      \put(206,87){\scriptsize $<P_{Tin}^2>(GeV/c)$}
      \put(276,120){\rotatebox{90} {\scriptsize $1/N_{ev}\cdot dN_{ch}/dP^2_T$}}
      \put(300,87){\scriptsize $<P_{Tout}^2>(GeV/c)$}
   \end{picture}
\end{figure}

\section{Comparison with experiment data}

\begin{figure}[htb]
\centerline{\epsfig{file=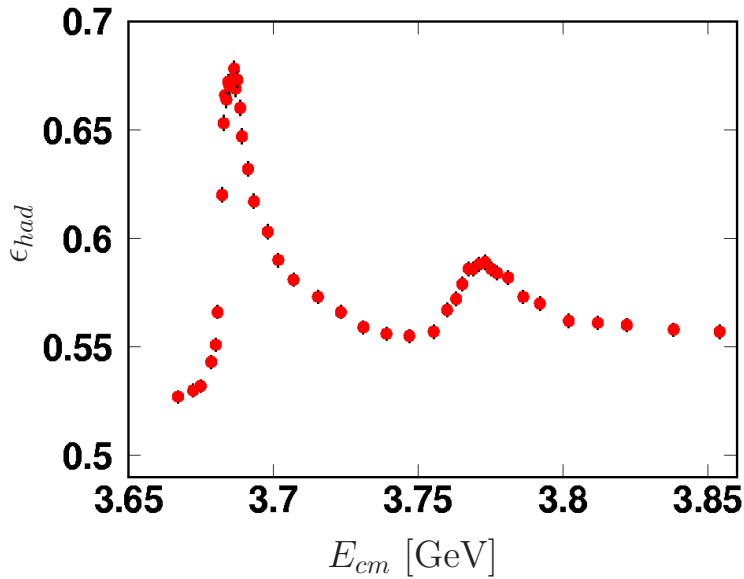,width=10cm,angle=0}}
\caption{The efficiencies for detection of the inclusive hadronic events at
each nominal center-of-mass energies.} 
\label{eff}
   \begin{picture}(5,5)
      \put(90,160){\rotatebox{90} {\large $\epsilon_{had}$}}
      \put(210,45){\large $E_{cm}$ [GeV]}
   \end{picture}
\end{figure}

Properties of hadronic events can be described by distributions of their
kinematic variables. Fig.~\ref{psi2} show the distributions of kinematic
variables of hadronic events produced at $\sqrt{s}=3.773$ GeV. In the
figure, shadows show the distributions of the variables from
data~\cite{chenjc_talk}, and lines represent the distributions from 
Monte Carlo sample generated with this generator, in which the detector
simulation is based on GEANT3 package~\cite{simbes}. 
In Fig.~\ref{psi2}, a) is the charged 
multiplicity $N_{ch}$ of events; b) is the $KNO$ scaling distribution~\cite{KNO};
c), d) and e) are the thrust distribution~\cite{thrust}, the sphericity
distribution and aplanarity distribution of events~\cite{sph}, respectively;
f) is the 
distribution of the Feynman scaling variable. Kinematic properties of 
hadronic events are described by g) pseudo-rapidity , and 
transverse momentum distributions h), n) and i).
From Fig.~\ref{psi2} one
can see that the qualities of the general simulation to the data are
satisfied.

\section{Application of the generator}
Feeding the reconstructed Monte Carlo events for $e^+e^-\rightarrow hadrons$
generated at each of the nominal center-of-mass energy points, at which
the cross section scans were performed, into the analysis
program to select the inclusive hadronic events one can determine the
detection efficiencies $\epsilon_{had}(s)$. Fig.~\ref{eff} shows the
Monte Carlo efficiencies for
detection of the hadronic events produced at the different nominal
center-of-mass energies from 3.650 to 3.872 GeV. From the figure one can
see that the efficiencies at the different energies with different physics
processes involved are quite different. The efficiencies determined with
the generator were used in
the measurement of $R$ values around 3.650 and at 3.773 GeV~\cite{R_uds}
which were used to
extract the inclusive branching fractions of $\psi(3770)$
decay to $D\overline{D}$, and
in the cross section scan experiment in the energy region covering both
of $\psi(3686)$ and $\psi(3770)$ resonances performed with BES-II detector
at the BEPC collider in Beijing~\cite{scan_1}.
For more precise measurements of the resonance parameters obtained by
fitting the cross sections for inclusive hadron production
in the energy range, where the complicate physical structure like
the $\psi(3686)$, $\psi(3770)$ and $\psi(4040)$ resonances involved, the correct
determination of the detection efficiencies is crucial. 

\section{Conclusion}
A new generator with $\alpha^2$ order radiative correction for the 
full simulation of $e^+e^-\rightarrow hadron$ process in
the cross section scan 
experiment in the energy region covering multiple resonance structure
is present. The complex resonances structure 
such as the $\psi(3770)$, $\psi(3686)$,
$J/\psi$, $\psi(4040)$ and $\psi(4160)$ as well as the continuum
QED production and the resonances with masses below 2 GeV/$c^2$, are all
included in the
generator. The generator reproduces the 
properties of hadronic events very well and the detector efficiency
can be determined conveniently for the hadronic cross section scan
experiment.

\section{Acknowledgment}
This work is supported in part by the National Natural Science Foundation
of China under contracts No. 10491304.

\end{document}